\documentclass[12pt]{iopart}
\usepackage{iopams}
\usepackage{bm}

\begin{document}

\title[ Multichannel Coulomb scattering problem]{Multichannel Coulomb Scattering 
	with Asymptotic Non-adiabatic Coupling }

\author{S L Yakovlev$^1$
N Elander$^2$
}

\address{${^1}$ Department of Computational Physics, St Petersburg State University, 199034, St. Petersburg, Russia}
\address{${^2}$ Department of Physics, Stockholm University,  Alba Nova University Center, SE 106 91, Stockholm, Sweden}

\ead{s.yakovlev@spbu.ru} 
\ead{elander@physto.se} 

\begin{abstract}
  The multi-channel Coulomb  scattering problem  in the adiabatic representation is considered.
  The non-adiabatic coupling matrix is assumed to have
  a {\bf non-zero} asymptotic behavior at large internuclear separations. The asymptotic solutions at large inter-nuclear distances are {therefore} constructed. The asymptotic boundary conditions with  S-matrix and K-matrix for the scattering problem are formulated on the basis of constructed asymptotic solutions.   
  
\end{abstract}


\submitto{\PS}

\maketitle

\section{Introduction}
\subsection{Background}

The adiabatic approach is one of the widely used methods for  theoretical study of a low-energy quantum collision problem in atomic and molecular physics \cite{MM}. In the context of the two-atom collision problem, this consists of using the eigenfunctions of the electronic sub-Hamiltonian as the basis for expanding the total wave function, which then leads to the resulting multichannel equations.

The main approximations within this approach includes  the Born-Oppenheimer approximation \cite{BO}, which corresponds to neglecting all  non-adiabatic couplings.
Without approximations the adiabatic representation is complete. All  couplings are kept in the formalism in this case and, therefore, solving the problem one obtains   
the exact solution to the Schr\"odinger equation in the adiabatic representation.
The main, and computationally challenging, difficulty of practical solving the scattering problem in the adiabatic representation is related to the non-adiabatic couplings between the channels.
Critically, these may happen
in the region of an {avoided} crossing of electronic energy levels which, as a rule, appears at finite inter-atomic distances.
Conventionally, one can overcome { this} difficulty  by  reformulating  the problem into the diabatic representation or into the more flexible  split diabatic representation (see, for example, \cite{ThrularandMead,Esry} and references therein). 
Another source of non-trivial
non-adiabatic couplings is associated
with the so-called molecular-state problem (sometimes also called the electronic translation problem), which appears in the formalism as the non-zero limit of the non-adiabatic coupling matrix elements when the internuclear distance tends  to infinity
\cite{DelosandThorson1979}-\cite{Belyaev2001}. For this kind of coupling the transformation of the equations into the diabatic form does not solve the problem.   
It is well understood \cite{Belyaev1999}-\cite{Belyaev2010} that the physical reason for these non-vanishing asymptotic couplings is due to the fact that the adiabatic approach is based on the molecular representation and, hence, on the molecular coordinates, typically, the Jacobi molecular coordinates, in which electrons are measured from the centre of nuclear mass.
The problem is that the molecular coordinates which are used to describe fixed-nuclei molecular states of the collision complex at small and intermediate distances are not suited for the description of the free atoms in the asymptotic region.
The difference arises from the fact that the inter-atomic vectors connecting the centres of mass of colliding atoms do not coincide with the internuclear vectors which connect the centers of mass of the respective nuclei.
This results in non-zero asymptotic couplings in coupled channel equations calculated in the molecular representation.
Thus, non-zero asymptotic couplings in coupled channel equations are fundamental features of the standard adiabatic approach.
This property, if present, {implies that} the simple Born-Oppenheimer asymptotic form of the scattering wave function in the adiabatic representation 
{is no longer valid}.

One of the
successful  methods for constructing a suitable asymptotic form of the wave function in the adiabatic representation is provided by the re-projection procedure 
\cite{Belyaev1999}-\cite{Belyaev2010}. 
This procedure utilizes the physically motivated asymptotic form of the total wave function which is represented in the channel specific Jacobi coordinates by the atomic eigenstates of non-interacting atoms. The asymptotic form of the wave function in the adiabatic representation is then calculated by re-projecting this total asymptotic wave function onto the adiabatic (molecular) basis. 

Asymptotic form of the solution in the special case of three-body scattering problem in the adiabatic representation was studied in \cite{PV1979}-\cite{PSV1981} and \cite{Melezhik1900}. One of the first attempts for direct constructing of the asymptotic solution for a special form of coupled equations with non-adiabatic coupling was made in \cite{Korobov}.  This method had found then wide applications in calculations of the meso-atomic and meso-molecular processes \cite{PonomarevCo}.  

In the recent paper \cite{Yakovlev2018},  the leading terms of asymptotic solutions of coupled equations were investigated in details. {Based} on the fact that the complete adiabatic representation automatically generates the correct asymptotic form of the wave function in the adiabatic representation, 
the asymptotic wave function is calculated in \cite{Yakovlev2018} from the asymptotic solutions of the adiabatic coupled channel equations in the region of large internuclear distances. {This} was done for the very leading order of the multichannel equations by neglecting the centrifugal terms in order to stay within the same conditions  as in the re-projection formalism. In this case the results of \cite{Yakovlev2018} coincide  with the re-projection method results \cite{Belyaev1999}-\cite{Belyaev2010}. The neutral atoms  in the asymptotic configuration  was another attribute of those papers. 
 In the present paper we generalize our approach in such a way that it is capable { to take  the centrifugal term and the asymptotic Coulomb interaction between colliding atoms, which has critical importance for practical solution of scattering problems for atoms and ions explicitly, into account.}

The paper is organized as follows. The mathematical definition of the problem is given in the following subsection. In section two we consider the general situation when the asymptotic non-adiabatic coupling matrix couples an arbitrary 
number of channels $N$.
In section three  the formulation of the asymptotic boundary conditions in two forms 
containing S-matrix and K-matrix is given. 
The last fourth section  concludes the paper.

\subsection{Multichannel adiabatic equations and 
asymptotic states}\label{sub1}
We consider the scattering problem for  the set of equations
\begin{eqnarray}
\hspace{-6mm}
\left[-\frac{d^2}{dr^2}+\frac{\ell(\ell+1)}{r^2}+\frac{C_j}{r}+{V}_j(r)-E\right]\!&F_j(r)= \nonumber\\
&\sum_{n\ge 1}\left[2{P}_{jn}(r)\frac{d}{dr}+{W}_{jn}(r)\right]\!F_n(r),
\label{eq1}
\end{eqnarray}
defined for integer $\ell\ge 0$ and real non-negative $r$, $0\le r <\infty$. {  $C_j$ represents the charge of the interacting particles.} The coefficients $V_j(r)$, $P_{jn}(r)$ and $W_{jn}(r)$ will be defined below.  

In order to correctly formulate the problem, the set (\ref{eq1}) has to be supplied by 
appropriate boundary conditions. 
While the boundary condition at $r=0$ is natural: $F_j(0)=0$, the asymptotic  form of $F_j(r)$  as $r\to\infty$ is not trivial and depends  on the decay rate of the coefficients 
$V_j(r)$, $P_{jn}(r)$ and $W_{jn}(r)$  when $r\to\infty$. The study of this asymptotic condition as $r\to\infty$ is the main task of this paper.

The set of equations (\ref{eq1}) conventionally represents the Schr\"odinger equation  for a system of two atoms by using the so-called adiabatic expansion \cite{MM} for the total wave function $\Psi$ of the system. The molecular basis set $\phi_k$ generates the expansion of the total wave function
\begin{equation}
\Psi(\bm{r},\bm{\xi})=Y_\ell^{m_{\ell}}({\hat r})\sum_{n\ge 1} \frac{F_{n}(r)}{r}\phi_n(\bm{\xi},r),
\label{expansion}
\end{equation}
where $Y_\ell^{m_\ell}$ are  the standard spherical functions, for which $\ell$ and $m_\ell$ denote the total angular momentum quantum numbers, $\bm{r}$ is the internuclear relative position vector, the unit vector $\hat r$ is given by $\bm{r}/r$, where $r=|\bm{r}|$,
{while $\bm{\xi}$ represents the electronic degrees of freedom.}
The above expansion  assumes  the following form of the two-atomic Hamiltonian
\begin{equation}
H= -\frac{\hbar^2}{2M} \Delta_{\bm{r}}
+ h(\bm{r},\bm{\xi}),
\label{Hamiltonian}
\end{equation}
with 
$M$ denoting the reduced  mass {of the two interacting nuclei}. The sub-Hamiltonian $h(\bm{r},\bm{\xi})$ containing all interaction potentials  governs the dynamics of electrons in the field of the ``frosen" nuclei. The molecular basis set $\phi_n(\bm{\xi},r)$ is then formed by eigenfunctions of the Hamiltonian  $h(\bm{r},\bm{\xi})$
\begin{equation}
h(\bm{r},\bm{\xi}) \phi_n(\bm{\xi},r)=\lambda_n(r)\phi_n(\bm{\xi},r).
\label{phi-base}
\end{equation}
The eigenfunctions obey the orthonormality and completeness conditions
\begin{equation}\label{ortho}
\langle \phi_j|\phi_n\rangle=\delta_{jn}, \ \ \sum_{n\ge 1}|\phi_n(\bm{\xi},r)\rangle\langle \phi_n(\bm{\xi}',r|=
\delta(\bm{\xi}-\bm{\xi}'),
\end{equation}
where $\delta_{jn}$ is the Kroneker symbol and $\delta(\bm{\xi}-\bm{\xi}')$ is the delta-function.
The relative internuclear position vector $\bm{r}$ { here serves}  as a parameter.
{ In applications it is quite common }
that the eigenfunctions $\phi_k(\bm{\xi},r)$ and the eigenvalues $\lambda_k(r)$ depend on the magnitude $r=|\bm{r}|$ of the vector $\bm{r}$.
This property corresponds \cite{MaciasRiera1982}, for example, to the case when only molecular $\Sigma$ states contribute to the problem.
These assumption can also be adapted to a more general case, which involves some additional small complications to the coupling matrices
\cite{Belyaev2001}.
One more important property of the molecular basis is  that  $\phi_j$ can always be chosen real since the sub-Hamiltonian $h(\bm{r},\bm{\xi})$ is Hermitian. This then leads to the anti-symmetry of the coupling matrix $P_{jm}(r)$.  

The matrices in  equations (\ref{eq1}) can now be expressed in terms of the molecular basis by following equations:
\begin{eqnarray}
V_j(r)+C_j/r=\frac{2M}{\hbar^2}\lambda_j(r) \label{V}, \\
P_{jn}(r)=\frac{2M}{\hbar^2}\langle \phi_j|\frac{\partial}{\partial r}|\phi_n\rangle \label{A}, \ \
W_{jn}(r)=\frac{2M}{\hbar^2} \langle \phi_j|\frac{\partial^2}{\partial r^2}|\phi_n\rangle \label{W}.
\end{eqnarray}
Here, the brackets in matrix elements mean the integration over electronic degrees of freedom $\bm{\xi}$.
The completeness of the basis leads to the relationship  for matrices $P_{jn}(r)$ and $W_{jn}(r)$
\begin{equation}
W_{jn}(r)= \sum_{m\ge 1} P_{jm}(r)P_{mn}(r)+\frac{d P_{jn}(r)}{dr}.
\label{W-A}
\end{equation}
Since the eigenfunctions $\phi_j$ are real, the matrix $P_{jn}(r)$ is antisymmetric, i.e. $P_{jn}(r)=-P_{nj}(r)$, and, hence, it is off-diagonal $P_{jj}(r)=0$.
The constant parameter $E$ in (\ref{eq1}) represents a reduced total energy in the  colliding system.

The asymptotic behavior of the potentials $V_j(r)$ and the matrices $P_{jn}(r)$ 
is
essential for the asymptotic analysis of the solution to the equation  (\ref{eq1}).
The asymptote of $V_j(r)$ determines the asymptotic thresholds $\epsilon_j$ 
\begin{equation}\label{thresholds}
\epsilon_j = \lim_{r\to\infty} V_j(r).
\end{equation}
In this paper we assume that the potentials $V_j(r)$ are short-range, that means
\begin{equation}\label{V-sr}
\lim_{r\to\infty} r^{2+\delta}[V_j(r)-\epsilon_j]=0
\end{equation}
for  $\delta>0$. The matrix $P_{jn}(r)$ determines the asymptotic couplings $a_{jn}$ through the formula
\begin{equation}\label{couplings}
a_{jn} = \lim_{r\to\infty} P_{jn}(r).
\end{equation}
As for potentials, it is required
\begin{equation}\label{P-sr}
\lim_{r\to\infty} r^{2+\delta}[P_{jn}(r)-a_{jn}]=0
\end{equation}
for  $\delta>0$. These ``short-range" conditions lead to the following asymptotic properties of the matrices in the equations (\ref{eq1})
\begin{eqnarray}
V_j(r)\sim \epsilon_j +O(r^{-(2+\delta)}), \label{V}\\
P_{jn}(r) \sim a_{jn} +O(r^{-(2+\delta)}), \label{P}\\
W_{jn}(r) \sim \sum_{m\ge 1} a_{jm}a_{mn} + O(r^{-(2+\delta)}) \label{W}.
\end{eqnarray}
From these properties, {\bf equation} (\ref{eq1}) can be 
recast into the form
\begin{eqnarray}
\left[-\frac{d^2}{dr^2}+\frac{\ell(\ell+1)}{r^2}+\frac{2k_j\eta_j}{r}-k^2_j
\right]F_j(r)= \nonumber\\ 
\sum_{n\ge 1}\left[2\,{a}_{jn}\frac{d}{dr}+ \sum_{m\ge 1} a_{jm}a_{mn}  \right]F_n(r)+ \sum_{m\ge 1} Q_{jm}(r)F_m(r)
,
\label{eq1-as}
\end{eqnarray}
where $k^2_j=E-\epsilon_j$ and $\eta_j=C_j/(2k_j)$.
In the last equation, couplings of the order $O(r^{-(2+\delta)})$ and less have been denoted by $Q_{jm}(r)$. In applications, as a rule, the remainders in (\ref{V}-\ref{W}) vanish as $r\to \infty$ much faster then in those formulas, but for the generality we leave ourselves with exposed decay rate, noticing that this already allows us to consider the Coulomb and centrifugal interactions to be the leading non-trivial ones.   

It follows from  the formal scattering theory \cite{Newton} and form the theory of asymptotic expansions for ordinary differential equations \cite{Wasow}  that terms of the order $O(r^{-(2+\delta)})$ do not affect the asymptotic
behavior of the solution terms of the orders up to $O(r^{-2})$ as $r\to \infty$. Therefore, in the current problem, studying the asymptotics of the solution to (\ref{eq1-as}) by setting $Q_{jm}(r)=0$  gives the correct asymptotic form of leading terms  of the solution to the equation
(\ref{eq1}).

Depending on the values of the asymptotic couplings $a_{jn}$, two cases should be distinguished: ({\it i}) $a_{jn}=0$ for all $j,n$; and
({\it ii}) there exists {
 a number} $N$ such that $a_{jn}\ne 0$ for $j,n \le N$ and $a_{jn}=0$, if $j>N$ or $n>N$.
The former case ({\it i}) is somewhat ``conventional" and corresponds to the Born-Oppenheimer type of asymptotic states \cite{BelyaevPS}.
As such, due to (\ref{V-sr}) and (\ref{P-sr}), the set of equations given in (\ref{eq1-as}) becomes decoupled in the limit $r\to\infty$, and  takes the form
\begin{eqnarray}
\left[-\frac{d^2}{dr^2}+\frac{\ell(\ell+1)}{r^2}+\frac{2k_j\eta_j}{r}-k^2_j
\right]F_j(r)= 0.
\label{eq2}
\end{eqnarray}
The two linearly independent solutions of (\ref{eq2}) are given by Coulomb functions $H^\pm_\ell(\eta_j,k_jr)= G_\ell(\eta_j,k_jr)\pm i F_\ell(\eta_j,k_jr)$ \cite{abram}
\begin{equation}\label{BO+-}
\hspace{-5mm}
F^{\pm}_j(r,k_j)= H^\pm_\ell(\eta_j,k_jr)\sim \exp\{\pm i[ k_jr-\eta_j\log(2k_jr)+\sigma^j_\ell - \frac{\pi\ell}{2}]\},
\end{equation}
where the channel momenta $k_j$ are given by
\begin{equation}
\label{k}
k_j=\sqrt{E-\epsilon_j}\ge 0.
\end{equation}
These solutions  provide us with the basis for the asymptotic form of the solution to the equation (\ref{eq1}) as $r\to\infty$
\begin{equation}\label{F-as}
F_j(r) \sim k^{-{1/2}}_j[\,b^+_j F^+_j(r,k_j) + b^-_jF^-_j(r,k_j)\,].
\end{equation}
The factor $k^{-1/2}_j$ is inserted in order to have the proper normalization in (\ref{F-as}) for incoming and outgoing waves on the unite flux.    
The scattering matrix can now be  defined as the transformation between incoming and outgoing amplitudes
\begin{equation}\label{S-BO}
-b^+_j=\sum_{n}S_{jn}b^-_n.
\end{equation}
If the leading terms of the solutions $F^{\pm}_j(r,k_j)$ from (\ref{BO+-}) are only taken into account in (\ref{F-as}), then this formula
is transformed into
\begin{equation}
\hspace{-18mm}
F_j(r) \sim e^{i\frac{\pi\ell}{2}}k^{-{1/2}}_j
[\,d^-_j \exp\{-i(k_ir-\eta\log2k_jr)\} + d^+_j \exp\{i(k_jr-\eta_j\log 2k_jr)\}].
\label{Fas-Coul}
\end{equation}
The $S$-matrix definition (\ref{S-BO}) in this case takes the form 
\begin{equation}
(-1)^{\ell+1}d^+_j=\sum_{m} e^{i\sigma^j_\ell} S_{jm} e^{i\sigma^m_\ell}d^-_m.
\label{s-matr-Coul}
\end{equation}

The latter case ({\it ii}) is significantly more complicated, and is studied in the subsequent sections.

\section{General case of multichannel equations}
In this section we consider the general situation, i.e. the case ({\it ii}) described in subsection \ref{sub1}, when $N$ states remain asymptotically coupled as $r\to \infty$.
Here, the set of equations (\ref{eq1}), (\ref{eq1-as}) is asymptotically split into two pieces:
the nontrivial $N\times N $ system for components $F_j$, $j=1, \ldots, N$; and the trivial decoupled set of the form (\ref{eq2})
for components $F_j$, $j>N$.
The latter leads to the asymptotic form of the components $F_j$, $j>N$,  given in Eq.~(\ref{F-as}).
The asymptotic form of the former components for $j=1,\ldots, N$ should be constructed by solving the relevant set of equations with non-trivial asymptotic coupling. As it was argued  in subsection \ref{sub1}, this set is obtained from (\ref{eq1}) or (\ref{eq1-as}) by neglecting all terms of the order $O(r^{-2-\delta})$ and less. The resulting asymptotic $N\times N$ set of equations  
 is
given by:
\begin{eqnarray}
\hspace{-15mm}\left[-\frac{d^2}{dr^2}+\frac{\ell(\ell+1)}{r^2}+\frac{2k_j\eta_j}{r}-k^2_j\right]&\Phi_j(r)= \nonumber\\
&\sum_{n\ge 1}\left[2\,a{\hat a}_{jn}\frac{d}{dr}+ a^2\sum_{m\ge 1} {\hat a}_{jm}{\hat a_{mn}}  \right]\Phi_n(r)
.
\label{eq2-as}
\end{eqnarray}
Here for the convenience we have explicitly introduced  the coupling constant $a$ by the definition
\begin{equation}
a_{jm} \equiv a {\hat a}_{jm}.
\label{a}
\end{equation}

In applications for atom-atom collisions  the coupling is small, i.e. $|a|\ll 1$, since it is proportional to $\sqrt{m/M}$, where $m$ is the electron mass and $M$ is the nucleus mass. In view of that we will solve the set (\ref{eq2-as}) perturbatively.
In the case of small coupling constant $a$ the solution to (\ref{eq2-as}) can be represented as
\begin{equation}
\Phi_j(r)=G^{(0)}_j(r) + a G^{(1)}_j(r) +O(a^2).
\label{Fa}
\end{equation}
For $G^{(0)}_j(r)$ we obtain the homogeneous equation
\begin{eqnarray}
\hspace{-15mm}\left[-\frac{d^2}{dr^2}+\frac{\ell(\ell+1)}{r^2}+\frac{2k_j\eta_j}{r}-k^2_j
\right]&G^{(0)}_j(r)= 0
.
\label{eq3-as}
\end{eqnarray}
The $N$ linearly independent solutions to (\ref{eq3-as}) are given by the Coulomb waves
\begin{equation}\label{Coul-w+-}
G^{(0)n\pm}_j(k_n,r)= \delta_{jn}H^\pm_\ell(\eta_n,k_nr)
\ \
n=1,...,N, 
\end{equation}
i.e., $N$ solutions for the sign $+$ and $N$ solutions for the sign $-$.
For $G^{(1)}_j(r)$ we obtain the inhomogeneous equation
\begin{eqnarray}
\hspace{-15mm}\left[-\frac{d^2}{dr^2}+\frac{\ell(\ell+1)}{r^2}+\frac{2k_j\eta_j}{r}-k^2_j
\right]&G^{(1)}_j(r)=
\nonumber\\
& 2\,{\hat a}_{jn}\frac{d}{dr}H^\pm_\ell(\eta_n,k_nr)
.
\label{eq3-as}
\end{eqnarray}
The asymptotic  solution to this equations when $r\to\infty$ can be constructed if the following asymptotic representation for derivative of $H^\pm_\ell$ is used, which can be derived from asymptotic representations for Coulomb wave functions \cite{abram}
\begin{equation}
\hspace{-12mm}\frac{d}{d\rho}H^{\pm}_\ell(\eta,\rho)  =
\pm i (1-\frac{\eta}{\rho})H^{\pm}_\ell(\eta,\rho)\mp i \frac{\ell(\ell+1)+\eta^2 \mp i\eta}{2\rho^2}H^{\pm}_\ell(\eta,\rho)+O(\rho^{-3}).
\label{H-as}
\end{equation}
The  asymptotic solution which obeys the equation (\ref{eq3-as}) up to the terms 
of the order $O(r^{-2})$ is represented
by the following expression
\begin{equation}
\hspace{-12mm}G^{(1)n\pm}_j(k_n,r)=
(\tau^{(0)\pm}_{jn}+\frac{\tau^{(1)\pm}_{jn}}{r}+\frac{\tau^{(2)\pm}_{jn}}{r^2}) H^\pm_\ell(\eta_n,k_nr)  + O(r^{-3}).
\label{F1}
\end{equation}
The coefficients $\tau^{(p)\pm}_{jn}$ for $j\ne n$ can now be calculated by introducing (\ref{H-as}) and  (\ref{F1}) into (\ref{eq3-as}) and then by  equating terms with equal degrees of $r^{-1}$. The results are given by
\begin{eqnarray}
\tau^{(0)\pm}_{jn}=\mp2i\frac{k_n}{k^2_j-k^2_n}{\hat a}_{jn}, \nonumber \\
\tau^{(1)\pm}_{jn}=2\frac{(k_j\eta_j-k_n\eta_n)\tau^{(0)\pm}_{jn} \pm i\eta_n{\hat a}_{jn}}{k^2_j-k^2_n}, \nonumber \\
\tau^{(2)\pm}_{jn}=2\frac{k_j\eta_j-k_n\eta_n \mp ik_n}{k^2_j-k^2_n}\tau^{(1)\pm}_{jn}
\pm i \frac{\ell(\ell+1)+\eta^2_n \mp i \eta_n}{k_n(k^2_j-k^2_n)}{\hat a}_{jn}. \nonumber \\
\label{tau}
\end{eqnarray}
For $j=n$ we have $\tau^{(p)\pm}_{jj}=0$.
Finally, the asymptotic solution $\Phi^{n\pm}_j$ is represented as the sum of two terms
\begin{equation}
\Phi^{n\pm}_j(k_n,r) =G^{(0)n\pm}_j(k_n,r)+ a G^{(1)n\pm}_j(k_n,r).
\label{Phi-n+-}
\end{equation}

The components $\Phi^{n\pm}_j$ 
for $j=1,\ldots,N$ together with components $F^\pm_j$ from (\ref{BO+-}) for $j>N$
form the basis for the asymptotic boundary conditions for the components $F_j(r)$ of the solution to the equations (\ref{eq1})
\begin{eqnarray}
F_j(r)\sim  \sum_{n=1}^{N}
{k_n}^{-1/2}\left[\,b^-_n \Phi^{n-}_j(k_n,r) + b^{+}_n\Phi^{n+}_j(k_n,r)\,\right], \ \ j=1,\ldots,N, \label{F-j-1} \\
F_j(r)\sim {k_j}^{-1/2}\left[\, b^-_j F^-_j(r,k_j) + b^+_j F^+_j(r,k_j) \, \right], \ \ j> N .
\label{F-j-2}
\end{eqnarray}
These boundary conditions are our main result that completes  the formulation of  the scattering problem
for the set of equations (\ref{eq1}) in the general situation.
The respective $S$-matrix is then defined as the transformation matrix between the incoming and outgoing amplitudes in scattering channels
\begin{equation}\label{S-Nmod}
-b^+_j=\sum_{n\ge 1}S_{jn}b^-_n.
\end{equation}

The derived formulas (\ref{F-j-1}) and (\ref{F-j-2}) give the leading with respect to the coupling constant $a$ terms.
Subsequent terms of the decompositions can be obtained (if it is necessary) by implementing the standard prescription of the perturbation theory \cite{Landau}. 

It is useful to give the expression for $\Phi^{n\pm}$ in the leading order with respect to $r\to\infty$
\begin{equation}
\Phi^{n\pm}_j(k_n,r)\sim t^\pm_{jn}H^{\pm}_{\ell}(\eta_n,k_nr).
\label{Phi-as}
\end{equation}
Here $t^{\pm}_{jn}$ is given by the same expression as in the non-Coulomb case \cite{Yakovlev2018}
\begin{equation}
t^{\pm}_{jn}=\delta_{jn} \mp \frac{2ik_n}{k^2_j-k^2_n}a_{jn}(1-\delta_{jn}).
\label{t}
\end{equation}

At the end of this section let us give the respective  formulas for the non-Coulomb case, which follow immediately from (\ref{Coul-w+-}), (\ref{F1}) and (\ref{tau}) when
all $\eta_j=0$, $j=1,...,N$
\begin{equation}
G^{(0)n\pm}_j(k_n,r)= h^{\pm}_{\ell}(k_nr)\delta_{jn},
\label{F00}
\end{equation}
where $h^\pm_\ell$ are the Riccati-Hankel functions.
\begin{equation}
G_j^{(1)n\pm}(k_n,r)=
(T^{(0)\pm}_{jn}+\frac{T^{(1)\pm}_{jn}}{r}+\frac{T^{(2)\pm}_{jn}}{r^2})(1-\delta_{jn}) h^\pm_\ell(k_nr)  + O(r^{-3}),
\label{F11}
\end{equation}
where $T^{(p)\pm}$ are given by the following expressions
\begin{eqnarray}
T^{(0)\pm}_{jn}=\mp2i\frac{k_n}{k^2_j-k^2_n}{\hat a}_{jn}, \nonumber \\
T^{(1)\pm}_{jn}=0, \nonumber \\
T^{(2)\pm}_{jn}=
\pm i \frac{\ell(\ell+1)}{k_n(k^2_i-k^2_n)}{\hat a}_{jn}. \nonumber \\
\label{TTT}
\end{eqnarray}

The asymptotic solution $\Phi^{n\pm}_j(k_n,r)$ is then given by the same formula (\ref{Phi-n+-}) as in the Coulomb case.


\section{S-matrix and K matrix}
In this section in addition to the $S$-matrix formulation (\ref{F-j-1}-\ref{S-Nmod}) we derive the asymptotic solution in terms of the $K$-matrix.
It is convenient to combine components  of the solutions $\Phi^{n\pm}_j(k_n,r)$ for $j=1,...,N$ and $F^{\pm}_j(k_j,r)$ in unified components as
${ U}^{n\pm}(r)$
\begin{eqnarray}
{U}^{n\pm}_j(k_n,r)=k^{-1/2}_n\Phi^{n\pm}_j(k_n,r), \ \ j=1,...,N, \\
{U}^{n\pm}_j(k_n,r) = \delta_{jn}k_n^{-1/2} F^{\pm}_n(k_n,r), \ \ j>N.
\label{U}
\end{eqnarray}
Let us also make a special choice for amplitudes $b^{-}_j$
\begin{equation}
b^{-}_j=\delta_{jm},
\label{b-minus}
\end{equation}
then   (\ref{F-j-1}-\ref{F-j-2}) take the following form
\begin{equation}
F_j^m(r)=U_j^{m-}(k_m,r)-\sum_{n} U_j^{n+}(k_n,r)S_{nm}.
\label{FF}
\end{equation}
The representation contains complex ingredients. In many respects it is convenient
work with real quantities. In order to derive the real solutions which are related to the reactance $K$ matrix we note first that $U_j^{n+}$ and $U_j^{n-}$ are complex conjugate, i.e.
$$
[U_j^{n+}(k_n,r)]^*=U_j^{n-}(k_n,r).
$$
Therefore, vectors with following components
\begin{equation}
\{{\cal B}_j(r)\}_n=  \frac{1}{2}[U_j^{n+}(k_n,r)+U_j^{n-}(k_n,r)],
\label{G}
\end{equation}
\begin{equation}
\{{\cal A}_j(r)\}_n=  \frac{1}{2i}[U_j^{n+}(k_n,r)-U_j^{n-}(k_n,r)]
\label{F}
\end{equation}
are real.  The Eq. (\ref{FF}) can now be rewritten with the  help of introduced quantities as
\begin{equation}
{\cal F}_j(r)=
\left\{
{\cal A}_j(r)-{\cal B}_j(r)\frac{1}{i}(I-S)(I+S)^{-1}\right\} \frac{-1}{4i}(I+S),
\label{KS}
\end{equation}
where ${\cal F}_j(r)$ is the vector with components $F^m_j(r)$ and $S$ and $I$ are matrices with matrix elements $S_{nm}$ and $\delta_{nm}$ respectively. In the
Eq. (\ref{KS}) the matrix multiplication from the right is assumed.
Let us now introduce the new asymptotic solution by the expression
\begin{equation}
{\cal G}_j(r) = {\cal A}_j(r)-{\cal B}_j(r)K, 
\label{KK}
\end{equation}
where
$K$ is the reactance matrix, which is defined via the $S$-matrix as
\begin{equation}
K= \frac{1}{i}(I-S)(I+S)^{-1}.
\label{KKSS}
\end{equation}
The inverse to (\ref{KKSS}) is given by Cayley transform and reads
\begin{equation}
S=(I+iK)^{-1}(I-iK)
\label{SSKK}
\end{equation}
representing the $S$-matrix in terms of the $K$-matrix.

\section{Conclusion}
In this research we have constructed the asymptotic solutions for multichannel equations with non-adiabatic coupling at large distances. The Coulomb interaction and centrifugal terms are taken into account explicitly. The exact representations are obtained for terms of the orders $O(1)$, $O(r^{-1})$ and $O{(r^{-2}})$ of the   asymptotic solution. With these asymptotic solutions the correct asymptotic boundary conditions for the scattering problem have been formulated. The $S$ and $K$ matrix are properly defined by the respective asymptotic representations of the solutions of the scattering problem.     
\ack
This work  is supported by the Russian Foundation for Basic Research grant No. 18-02-00492. The authors are thankful to A.K. Belyaev 
for discussions.



\Bibliography{99} 



\bibitem{MM} Mott N F and Massey H S W 1949 {\it The Theory of Atomic Collisions} (Oxford:Clarendon).
\bibitem{BO} Born M and Oppenheimer R 1927 {\it Ann. Phys.} (Leipzig) {\bf 87} 457.
\bibitem{ThrularandMead} Mead C A  and  Truhlar D G   1982 {\it J. Chem. Phys.} {\bf 77} 6090.
\bibitem{Esry} Esry B D and Sadeghpour H R 2003 {\it Phys. Rev.} A {\bf 68}  042706.
\bibitem{DelosandThorson1979}  Delos J B and Thorson W R 1979 {\it J. Chem. Phys.} {\bf 70} 1774.
\bibitem{Delos1981} Delos J B 1981 {\it  Rev. Mod. Phys. } {\bf 53} 287.
\bibitem{McCarroll1981} Gargaud M, Hanssen J, McCarroll R and Valiron P 1981 {\it J. Phys.} B {\bf 14} 2259.
\bibitem{MaciasRiera1982} Macias A and Riera A 1982 {\it Phys. Rep.} {\bf 90} 299.
\bibitem{Belyaev1999} Grosser J, Menzel T and Belyaev A K 1999 {\it Phys. Rev.} A {\bf 59 } 1309.
\bibitem{Belyaev2001} Belyaev A K, Egorova D, Grosser J  and Menzel T 2001 {\it Phys. Rev.} A {\bf 64} 052701.
\bibitem{Belyaev2002} Belyaev A K, Dalgarno A  and McCarroll R 2002 {\it J. Chem. Phys.} {\bf 116} 5395-5400.
\bibitem{Belyaev2010} Belyaev A K  2010 {\it Phys. Rev.} A {\bf 82} 060701(R).
\bibitem{PV1979} Ponomarev L I and Vinitsky S I  1979 {\it J. Phys.} B {\bf 12} 567.
\bibitem{PVV1980} Ponomarev L I, Vinitsky S I and Vukajlovic F R 1980 {\it J. Phys.} B {\bf 13} 847.
\bibitem{PSV1981} Ponomarev L I, Somov L N and Vukajlovic 1981 {\it J. Phys.} B {\bf 14} 591.
\bibitem{Melezhik1900}  Bracci L,  Chiccoli C, Fiorentini G,  Melezhik V S,  Pasini P and Wozniak J 1990
{\it Il Nuovo Cimento} {\bf 105} 459.


\bibitem{Korobov} Korobov V I  1994 {\it  J.Phys. B: At. Mol. Opt. Phys.} {\bf 27} 733.   
\bibitem{PonomarevCo} Adamczak A,  Faifman M P,  Korobov V I,  Melezhik V S,  Ponomarev L I,  Siegel R T, and  Wozniak J  1996
{\it Atomic Data and Nuclear Data Tables}  {\bf 62} (2) 255-344. 
\bibitem{Yakovlev2018} Yakovlev S L, Yarevsky E A, Elander N O and Belyaev A K  
2018 {\it Theor. Math. Phys.} {\bf 195}(3): 874-885.

\bibitem{Newton}Newton R G 1982 {\it Scattering theory of Waves and Particles} (New-York: Springer-Verlag)
\bibitem{Wasow} Wasow W  1965 {\it Asymptotic expansions for ordinary differential equations} (New-York $\cdot$London$\cdot$Sydney: Wiley\&Sons.Inc.)


\bibitem{abram} Abramowitz M and Stegan I A (eds) 1986 {\it Handbook of Mathematical Functions} (New York: Dover)


\bibitem{BelyaevPS} Belyaev A K 2009 {\it Physica Scripta} {\bf 80} 048113.
\bibitem{Landau} Landau L D and Lifshitz E M 1965 {\it Quantum Mechanics (Volume 3 of A Course of Theoretical Physics)} (Pergamon Press).
\endbib
\end{document}